\documentclass[aps,prl,twocolumn,letterpaper,groupedaddress,10pt]{revtex4-1}
\usepackage{amssymb,amsthm,amsmath,amsfonts}
\usepackage{graphicx,ulem,mathptmx}
\usepackage[pdftex,dvipsnames,usenames]{xcolor}
\usepackage[colorlinks=true,urlcolor=blue,citecolor=blue,linkcolor=blue]{hyperref}

\begin{document}

\title{Continuity of measurement outcomes}

\author{J. Sperling}\email{jan.sperling@physics.ox.ac.uk}
\affiliation{Clarendon Laboratory, University of Oxford, Parks Road, Oxford OX1 3PU, United Kingdom}

\date{\today}

\begin{abstract}
	It is demonstrated that the collapse of the wave function is equivalent to the continuity of measurement outcomes.
	The latter states that a second measurement has to result in the same outcome as the first measurement of the same observable for a vanishing time between both observations.
	In contrast to the exclusively quantum-physical collapse description, the equivalent continuity requirement also applies in classical physics, allowing for a comparison of both domains.
	In particular, it is found that quantum coherences are the single cause for measurable deviations in statistical properties due to the collapse.
	Therefore, the introduced approach renders it possible to characterize and quantify the unique features of the quantum-physical measurement problem within the framework of modern quantum resource theories and compare them to classical physics.
\end{abstract}

\maketitle

%%%%%%%%%%%%%%%%%%%%%%%%%%%%%%%%%%%%%%%%%%%%%%%%%%%%%%%%%%%%%%%%%%%%%%%%%%%%%%%%%%%%%%%%%%%%%%%%%%%%%%%%%%%%%%%%

	The collapse of the wave function (CWF) is a cornerstone of quantum physics and describes how a system responds to a measurement process \cite{H27,N32}.
	The consequence that the detection process can instantaneously alter the quantum state is a counterintuitive mechanism genuine to quantum physics.
	Despite having such a surprising property, the CWF makes predictions which have been repeatedly confirmed in experiments; see Ref. \cite{FTZWF15} for a recent implementation.
	In addition, the remote manipulation of quantum states, which is caused by the CWF, is a useful feature for novel applications, such as measurement-based quantum computation \cite{BBDRN09}.
	Still, the discontinuous reduction of the quantum state, also known as the measurement problem, is not fully understood yet.

	Independently from the CWF, the field of quantum coherence has been developed for the aim of studying and quantifying quantum phenomena as resources for practical applications \cite{SAP17}.
	Specifically, quantum superpositions have been identified as a unique foundation for quantum coherence \cite{A06,A14,SV15,WY16,TKEP17}.
	A comprehensive analysis of interferences and coherences in general theories can be found in Ref. \cite{BDW17}.
	Moreover, transformations between diverse notions of quantum coherence allow for a versatile utilization of different quantum effects, such as conversions between local nonclassicality and entanglement \cite{VS14,KSP16,CH16}.
	Furthermore, quantum correlations in bipartite systems---in combination with the influence of local measurements of one subsystem---have been studied and quantified, for instance, in relation to steering \cite{HF16} and quantum discord \cite{MYGVG16}; see also Ref. \cite{SRBL17}.
	Moreover, conditional quantum correlations in such scenarios have been experimentally investigated \cite{SBDBJDVW16,ASCBZV17}, and they connect to ancilla-assisted quantum protocols \cite{CSRBAL16,MZFL16}.
	These recent advances, using the projective properties of the CWF, hint at a deeper interconnection between application-based measures of quantum coherence and the fundamental aspect of the CWF \cite{YDXLS17}.

	The instantaneous CWF is inherent to quantum systems and, therefore, the formulation of a classical analog to the CWF appears to be an illusive undertaking, challenging classical intuitions about the physical world \cite{S35,W05}.
	While the Schr\"o\-dinger equation describes the continuous evolution, the collapse model represents a discontinuous alteration of the quantum state.
	For this reason, some theoretical approaches, with decoherence being arguably the most prominent example \cite{Z03}, attempt to replace the CWF with other continuous quantum mechanisms.
	Nevertheless, a classical correspondence is required to assess the quantumness of the measurement process, for instance, in the framework of quantum coherence, which is based on comparing quantum states with incoherent ones.
	This open problem of finding a classical counterpart to the CWF is addressed here; its resolution is shown to lead to new insights into a resource-theoretical interpretation of the measurement problem.

	In this contribution, it is demonstrated that the discontinuous CWF can be replaced by a continuity requirement, the continuity of measurement outcomes (CMO), by showing the equivalence of both concepts.
	Besides the resulting reinterpretation of the CWF as a consequence of a more intuitive and accessible principle, the CMO also applies in classical physics and can be related to conditional probabilities.
	Still, the implications of the CMO in the classical and quantum realm are distinctively different.
	This renders it possible to formulate measurable criteria, based on non-commuting observables, to verify the CWF through quantum coherences.
	Therefore, a substitution of the CWF with the equivalent CMO as the prime axiom for quantum measurements provides an experimentally accessible connection between fundamental aspects of the CWF and modern resource theories, relevant for applications of quantum information technology.

%%%%%%%%%%%%%%%%%%%%%%%%%%%%%%%%%%%%%%%%%%%%%%%%%%%%%%%%%%%%%%%%%%%%%%%%%%%%%%%%%%%%%%%%%%%%%%%%%%%%%%%%%%%%%%%%

	To formulate the desired equivalence, an observable with the spectral decomposition $\hat x=\sum_n x_n |x_n\rangle\langle x_n|$ is considered.
	To avoid that technical difficulties obstruct the physical meaning, it is assumed that this observable is non-degenerate and acts on a finite-dimensional Hilbert space.
	Born's rule states that the probability for the outcome $x_n$ is given by $P(x_n)=\langle x_n|\hat\rho|x_n\rangle$ when the system is in the quantum state $\hat\rho$ at the measurement time $t=0$.
	The collapsed state after this measurement is the eigenstate $\hat\rho'=|x_n\rangle\langle x_n|$ and further evolves according to the Schr\"odinger equation, labeled as $\hat\rho^{(t)}$ for $t>0$.

	One implication of the CWF is that a second measurement of $\hat x$ becomes deterministic when the waiting time after the first measurements tends to zero,
	\begin{align}
		\label{eq:CMO}
		\lim_{t\to 0^{+}}P^{(t)}(x_m|x_n)=\delta_{m,n},
	\end{align}
	where $t\to0^{+}$ indicates a limit to zero for positive delays ($t>0$) and $\delta$ denotes the Kronecker symbol.
	Here, $P^{(t)}(x_m|x_n)$ denotes the probability that $x_m$ is measured under the constraint that the first measurement outcome is $x_n$.
	Equation \eqref{eq:CMO} defines the CMO and states that the measurement outcomes of two subsequent measurements of the same observables becomes identical when the waiting time between the measurements approaches zero \cite{comment1}.
	The CMO is a result of the CWF as the collapsed state $\hat \rho'=\lim_{t\to0^{+}}\hat\rho^{(t)}=|x_n\rangle\langle x_n|$ implies $P^{'}(x_m|x_n)=\lim_{t\to0^{+}}P^{(t)}(x_m|x_n)=\langle x_m|\hat\rho'|x_m\rangle=\delta_{m,n}$.

	Interestingly, the inverse direction---the CMO [Eq. \eqref{eq:CMO}] implies the CWF---can be proven as well.
	Namely, the state $\hat\rho'$ instantaneously after the measurement can be expanded, in general, as $\hat\rho'=\sum_{k,l}\rho'_{k,l}|x_k\rangle\langle x_l|$.
	Then, the premise in Eq. \eqref{eq:CMO} implies that $\rho'_{n,n}=\langle x_n|\hat\rho'|x_n\rangle=1$ and $\rho'_{m,m}=0$ for $m\neq n$.
	Since $\hat\rho'$ is a positive semidefinite operator, this further implies that $\rho'_{k,l}=0$ for all $k\neq l$.
	Thus, the state after the first measurement has to have the expansion $\hat\rho'=|x_n\rangle\langle x_n|$, which is the one state postulated by the CWF model.

	Therefore, the CMO is demonstrated to be equivalent to the CWF.
	This equivalence enables one to start with the continuity requirement posed by the CMO as the primary axiom and degrade the discontinuous CWF to a conclusion from it, which is the opposite direction typically followed.
	One advantage is that it seems more natural to require that the measurement outcome is conserved for the same observable and a waiting time approaching to zero because there is no time for the system to evolve such that a different outcome is possible.
	In contrast to the clear interpretation of the CMO, the discontinuity of the instantaneous CWF is a more challenging concept.
	Another benefit is that Eq. \eqref{eq:CMO} makes sense in the context of classical measurements too.

	In classical statistical physics, the probability $P(X)$ of a system to be in a configuration in a set $X$ at a measurement time $t=0$ is described in terms of a classical probability distribution $P$.
	Specifically, the disjoint decomposition $\{X_m\}_m$ of the configuration space can be made such that the elements of $X_m$ lead to the outcome $x_m$ with the probability $P(X_m)$.
	After a waiting time $t$, the initial set $X$ is mapped to a new part of the configuration space, $X^{(t)}$.
	The probability that the system evolved into another set of configurations, $\tilde X$, is then given by the conditional probability $P(\tilde X|X^{(t)})=P(\tilde X\cap X^{(t)})/P(X^{(t)})$.
	As the continuous evolution implies $X=\lim_{t\to0^{+}} X^{(t)}$, it follows in the limit of zero delay for a second measurement that
	\begin{align}
		\label{eq:classicalCMO}
		\lim_{t\to 0^{+}}P(X_m|X_n^{(t)})
		=\delta_{n,m}
	\end{align}
	for the individual measurement outcomes.
	This is a conclusion of the definition of conditional probabilities, the identity $P(X_n\cap X_m)=P(X_n)$ for $n=m$, and $P(X_n\cap X_m)=P(\emptyset)=0$ for $m\neq n$.
	Moreover, relation \eqref{eq:classicalCMO}, reasoned by classical statistics, resembles the quantum version of the CMO given in Eq. \eqref{eq:CMO} and, thus, is the sought-after classical analog.

%%%%%%%%%%%%%%%%%%%%%%%%%%%%%%%%%%%%%%%%%%%%%%%%%%%%%%%%%%%%%%%%%%%%%%%%%%%%%%%%%%%%%%%%%%%%%%%%%%%%%%%%%%%%%%%%

	While the subsequent measurement of a single observable defines the CMO, it is known that the impact of th the CWF is most pronounced when combining two incompatible measurements, which, for example, manifests itself in Heisenberg's uncertainty relation \cite{H27}.
	Thus, the second measurement may be replaced by the observable $\hat y=\sum_{m}y_m|y_m\rangle\langle y_m|$, which does not commute with $\hat x$.
	Again, a vanishing waiting time between the measurements is considered, $t\to 0^{+}$.

	In the classical domain, the outcome $y_m$ is obtained for a subset $Y_m$ of the configuration space with a probability $P(Y_m)$ for the case of no prior measurement.
	If, however, a first measurement is conducted and yields the outcome $x_n$, then the conditional probability reads $P(Y_m|X_n)=P(Y_m\cap X_n)/P(X_n)$.
	Thus, the total probability to measure $y_m$ in this scenario is $P'(Y_m)=\sum_n P(X_m)P(Y_m|X_m)$.
	A highly relevant observation is that both expressions $P$ and $P'$ are identical in classical statistical physics as the law of total probability \cite{S95}, $P(Y)=P'(Y)$, is always satisfied.
	This raises the question if the same law holds true in quantum domain.

	If the first quantum measurement yields the outcome $x_n$, the quantum version of the CMO predicts the state $\hat\rho'=|x_n\rangle\langle x_n|$ directly after the first measurement.
	The conditional probabilities of the second measurement are consequently given by $P'(y_m|x_n)=\langle y_m|\hat\rho'|y_m\rangle=|\langle y_n|x_m\rangle|^2$.
	Thus, the total probabilities for the measurement outcomes of $\hat y$ read
	\begin{align}
		\label{eq:preXprob}
		P'(y_m)=\sum_{n} P(x_n)P'(y_m|x_n)=\langle y_m|\left(\sum_{n} P(x_n)|x_n\rangle\langle x_n|\right)|y_m\rangle.
	\end{align}
	By contrast, without the prior measurement of $\hat x$, the probabilities for measuring $\hat y$ are given by
	\begin{align}
		\label{eq:notXprob}
		P(y_m)=\langle y_m|\hat\rho|y_m\rangle.
	\end{align}
	Therefore, the probability distribution $P'(y_m)$ after the measurement of $\hat x$ differs from the distribution $P(y_m)$ obtained without a prior observation of $\hat x$ as long as the state exhibits quantum coherences, i.e., $\hat\rho\neq \sum_{n}P(x_n)|x_n\rangle\langle x_n|$ with $P(x_n)=\langle x_n|\hat\rho|x_n\rangle$, and the observables do not commute---implying a relation of the form $|\langle x_n|y_m\rangle|^2=\delta_{m,n}$ does not hold true.

	In the quantum measurement framework, the notion of incoherent states, which are of the general diagonal form $\sum_n p_n|x_n\rangle\langle x_n|$ for the basis $\{|x_n\rangle\}_n$ \cite{SAP17}, naturally occurs.
	In particular, the differences of the state $\hat\rho$ and the corresponding incoherent ensemble of collapsed states,
	\begin{align}
		\label{eq:ClassicalEnsemble}
		\hat\sigma=\sum_{n}|x_n\rangle\langle x_n|\hat\rho|x_n\rangle\langle x_n|=\sum_n P(x_n)|x_n\rangle\langle x_n|,
	\end{align}
	results in different probability distributions, $P(y_m)\neq P'(y_m)$, given in Eqs. \eqref{eq:notXprob} and \eqref{eq:preXprob}, respectively.
	In this context, it is also worth pointing out that Eq. \eqref{eq:ClassicalEnsemble} relates to the application of a so-called strictly incoherent operation \cite{WY16,YMGGV16}.

	In conclusion, the classical CMO implies that $P$ and $P'$ are identical, but the same does not hold true for quantum measurements.
	Consequently, the CWF has a measurable impact in the quantum domain and can be accessed through the notion of quantum coherence.
	The other way around, for any incoherent state, being invariant under CWF [$\hat\rho=\hat\sigma$], $P(y)=P'(y)$ holds true for all observables $\hat y$.
	Therefore, coherent states (with respect to the first measurement $\hat x$) are uniquely identified by their ability to produce an observable statistical difference, $P\neq P'$, for at least one second measurement $\hat y$.

	An interesting connection between the CWF and quantum coherence was previously reported in Ref. \cite{YDXLS17}.
	Here, however, it is important to stress that the presented results are obtained from a purely classical perspective on the CMO.
	Further, the implications valid in the classical framework (i.e., the law of total probabilities) have been shown to be violated by quantum physics.

%%%%%%%%%%%%%%%%%%%%%%%%%%%%%%%%%%%%%%%%%%%%%%%%%%%%%%%%%%%%%%%%%%%%%%%%%%%%%%%%%%%%%%%%%%%%%%%%%%%%%%%%%%%%%%%%

	In order to formulate an experimentally friendly criterion, the law of total variances \cite{S95} can be additionally considered.
	This relation from the theory of classical conditional probabilities reads
	\begin{align}
		\label{eq:totalVar}
		\mathbb V_{P(Y)}[Y]=\mathbb E_{P(X)}[\mathbb V_{P(Y|X)}[Y]]+\mathbb V_{P(X)}[\mathbb E_{P(Y|X)}[Y]],
	\end{align}
	where $\mathbb E_{P}$ and $\mathbb V_P$ denote the expectation value and variance for a distribution $P$, respectively.
	However, inserting the corresponding quantum-physical distributions, one readily gets
	\begin{align}
		\label{eq:VarPARTS}
		&\mathbb V_{P(y)}[y]=\langle (\Delta\hat y)^2\rangle_{\hat\rho}
		\text{ and}
		\\\nonumber
		&\mathbb E_{P(x)}[\mathbb V_{P'(y|x)}[y]]+\mathbb V_{P(x)}[\mathbb E_{P'(y|x)}[y]]
		= \mathbb V_{P'(y)}[y]=
		\langle (\Delta\hat y)^2\rangle_{\hat\sigma},
	\end{align}
	using the quantum-physical variances $\langle (\Delta\hat y)^2\rangle_{\hat\pi}=\langle \hat y^2\rangle_{\hat\pi}-\langle \hat y\rangle_{\hat\pi}^2$ for $\hat\pi\in\{\hat\rho,\hat\sigma\}$.
	Thus, if the quantum analog to the classical identity \eqref{eq:totalVar} is not satisfied,
	\begin{align}
		\label{eq:Condition}
		\mathbb E_{P(x)}[\mathbb V_{P'(y|x)}[y]]+\mathbb V_{P(x)}[\mathbb E_{P'(y|x)}[y]]-\mathbb V_{P(y)}[y]\neq0,
	\end{align}
	then the presence of coherence is certified through the consequences of the CWF.
	In that case, the measured quantum fluctuations for $\hat y$ are distinctively different depending on whether or not $\hat x$ was previously measured, $\langle (\Delta\hat y)^2\rangle_{\hat\rho}\neq\langle (\Delta\hat y)^2\rangle_{\hat\sigma}$.

%%%%%%%%%%%%%%%%%%%%%%%%%%%%%%%%%%%%%%%%%%%%%%%%%%%%%%%%%%%%%%%%%%%%%%%%%%%%%%%%%%%%%%%%%%%%%%%%%%%%%%%%%%%%%%%%

	Since the qubit is of fundamental importance as the basic unit of quantum information \cite{KL98,NC00}, such a system can be used to demonstrate the general function of the introduced methods.
	Suppose the two observables are given as
	\begin{align}
		\label{eq:exampleObservables}
		\hat x=\begin{bmatrix}
			-1 & 0 \\ 0 & 1
		\end{bmatrix}
		\text{ and }
		\hat y=\cos\vartheta\begin{bmatrix}
			-1 & 0 \\ 0 & 1
		\end{bmatrix}
		+\sin\vartheta\begin{bmatrix}
			0 & e^{-i\varphi} \\ e^{i\varphi} & 0
		\end{bmatrix},
	\end{align}
	which both have the possible outcomes $x,y\in\{+1,-1\}$.
	An arbitrary, mixed initial qubit state can be parametrized as
	\begin{align}
		\label{eq:exampleState}
		\hat\rho=\begin{bmatrix}
			1-p & \sqrt{p(1-p)}\gamma
			\\
			\sqrt{p(1-p)}\gamma^\ast & p
		\end{bmatrix},
	\end{align}
	for $p\in[0,1]$ and $|\gamma|\leq 1$.
	The probabilities for the measurement of $\hat x$ are $P(x=+1)=p$ and $P(x=-1)=1-p$, and non-zero off-diagonal elements ($p(1-p)|\gamma|^2\neq0$) define quantum coherences.
	From the spectral decomposition of $\hat y$, the conditional probabilities can be obtained, $P'(y=\pm 1|x=+1)=(1\pm\cos\vartheta)/2$ and $P'(y=\pm 1|x=-1)=(1\mp\cos\vartheta)/2$.

	Applying the law of total probabilities from classical statistics, cf. Eq. \eqref{eq:preXprob}, the probability distribution for the measurement of $\hat y$ after the measurement of $\hat x$ reads
	\begin{align}
		\label{eq:exampleClassicalP}
		P'(y=\pm 1)=\frac{1\pm(2p-1)\cos\vartheta}{2}.
	\end{align}
	In comparison, the probability distribution without the prior CWF can be put into the form
	\begin{align}
		P(y=\pm 1)
		=P'(y=\pm 1)
		\pm\mathrm{Re}(\gamma e^{i\varphi})\sqrt{p(1-p)}\sin\vartheta,
	\end{align}
	where the second summand accounts for the quantum interferences.
	Clearly, $P(y)$ is identical to $P'(y)$ [Eq. \eqref{eq:exampleClassicalP}] when the extra contributions vanish.
	This holds true iff $\sin\vartheta=0$, which is is true when $\hat x$ and $\hat y$ commute [cf. Eq. \eqref{eq:exampleObservables}], or $\mathrm{Re}(\gamma e^{i\varphi})\sqrt{p(1-p)}=0$, which is satisfied for any $\varphi$ if the state in Eq. \eqref{eq:exampleState} is incoherent (i.e., diagonal).
	Moreover, the trace-norm distance between the state $\hat \rho$ and its incoherent counterpart in Eq. \eqref{eq:ClassicalEnsemble}, $\hat\sigma=\left[\begin{smallmatrix}1-p&0\\0&p\end{smallmatrix} \right]$, can be obtained by varying over $\varphi$ and $\vartheta$ to probe all possible qubit observables (modulo the addition of the identity),
	\begin{align}
		\label{eq:TraceDistance}
		\|\hat\rho-\hat\sigma\|_{1}
		=\max_{\varphi,\vartheta}\sum_{y\in\{+1,-1\}}|P(y)-P'(y)|
		=2|\gamma|\sqrt{p(1-p)},
	\end{align}
	which is a quantifier of quantum coherence and, here, obtained from the incompatible implication of the principle of CMO in the classical and quantum domain.

	Furthermore, the condition in Eq. \eqref{eq:Condition} can be applied.
	For this purpose, the variances $\langle (\Delta\hat y)^2\rangle_{\hat\rho}$ and $\langle (\Delta\hat y)^2\rangle_{\hat\sigma}$ in Eq. \eqref{eq:VarPARTS} are computed, where the variance for $\hat\rho$ is obtained without a prior measurement of $\hat x$ and the second variance is obtained for the incoherent ensemble of collapsed states $\hat\sigma$ [Eq. \eqref{eq:ClassicalEnsemble}].
	For example, the difference in Eq. \eqref{eq:Condition} for the parameters $\varphi=-\arg\gamma$ and $\vartheta=\pi/2$ is
	\begin{align}
		\langle (\Delta\hat y)^2\rangle_{\hat\sigma}-\langle (\Delta\hat y)^2\rangle_{\hat\rho}
		= 4|\gamma|^2p(1-p),
	\end{align}
	which is the square of the trace distance in Eq. \eqref{eq:TraceDistance}.
	Consequently, the directly applicable criterion \eqref{eq:Condition} certifies the collapse of the qubit state as a result of its quantum coherences through measured fluctuations.
	It is worth stressing again that the applied condition is formulated in terms of variances that are necessarily identical in classical theories [Eq. \eqref{eq:totalVar}].

%%%%%%%%%%%%%%%%%%%%%%%%%%%%%%%%%%%%%%%%%%%%%%%%%%%%%%%%%%%%%%%%%%%%%%%%%%%%%%%%%%%%%%%%%%%%%%%%%%%%%%%%%%%%%%%%

	In summary, an equivalent paradigm to the CWF was found, the CMO.
	However, in contrast to the CWF, the CMO is directly applicable in classical statistics as well.
	Using conditional probabilities and the law of total probabilities, an analysis of the CMO was conducted, leading to two incompatible results in the quantum domain which should, however, be identical in classical physics.
	In contrast to a previous work \cite{YDXLS17}, here the concept of an incoherent state is derived from the analysis of the CMO in a purely classical framework.
	Specifically, incoherent state are uniquely obtained as those states which are indeed consistent with the classical prediction.
	Quantum coherence, on the other hand, leads to at least one collapse scenario which is incompatible with classical physics and verifiable with the derived technique.
	As an example, an experimentally accessible criterion was formulated in terms of variances.
	As a proof of concept, this method was then applied to an example to quantify coherence in connection with non-commuting measurements.

	In general, the aim of this work is to providing a deeper understanding of the instantaneous CWF from a classical perspective and characterizing this quantum phenomenon on a quantitative and measurable basis.

	Because the CMO is shown to be equivalent to the CWF, it is valid to confer to the CMO as the primary paradigm, whereby the CWF becomes the derived property.
	It is also noteworthy that the formulated equivalence combines the discontinuous collapse with a continuity requirement.
	To be clear, it is not proposed to replace the CWF with an alternative mechanism, which is the case for other approaches, e.g., in the context of decoherence \cite{Z03}.
	Rather, the CWF is substituted with the equivalent principle of the CMO.
	This change of perspective is preferable as the CMO has a correspondence in the classical domain, which cannot be simply found for the CWF.
	However, the classical and quantum versions of the CMO are only superficially identical.
	The resulting deviation renders it possible to discriminate the quantum measurement process from classical observations using quantum coherences, leading to another physical motivation of this resource which connects it to the measurement-induced collapse.
	This connection also achieves the desired quantitative assessment of the CWF.
	Specifically, this can be done by analyzing the uncertainties with and without the collapse of the state due to a prior measurement as derived in this work.

	In addition, the introduced concepts can be extended to more general scenarios.
	For instance, imperfections in the measurement process can have an impact on the collapsed state and are likely to soften the classical-quantum boundary.
	Furthermore, applying the proposed framework to a subsystem of a multipartite system will remotely influence the remaining parts, similarly to effects reported in previous studies, e.g., in Ref. \cite{SRBL17}.
	Thus, future investigations may establish collapse-based relations to nonlocal forms of quantum coherence, such as entanglement, in noisy environments.
	Also, the classical form of the CMO and the violation of the law of total probabilities in quantum physics hints at connections to somewhat related problems, such as contextuality and causality.

% 	A link between quantum coherence and the classical analog to the CWF is established here with the potential to inspire future research and which places the quantum phenomenon of the CWF within a resource-theoretic framework relevant for quantum technologies.

	Finally, it is also worth mentioning that the proposed change of perspective---from CWF to CMO---also redirects some interpretations of quantum physics, e.g., the seminal gedankenexperiment by Schr\"odinger \cite{S35}.
	That is, the important aspect of the CWF in the detection process from the CMO standpoint is that the first measurement of the state of the cat (say the outcome is ``dead'') is consistent with a second observation and does not alter the state to ``alive.''
	Or, in simple terms, the CMO implies that observations of ``zombie cats'' are excluded from both the classical and quantum realm.

%%%%%%%%%%%%%%%%%%%%%%%%%%%%%%%%%%%%%%%%%%%%%%%%%%%%%%%%%%%%%%%%%%%%%%%%%%%%%%%%%%%%%%%%%%%%%%%%%%%%%%%%%%%%%%%%

\begin{acknowledgments}
	This work has received funding from the European Union's Horizon 2020 Research and Innovation Program under grant agreement No. 665148 (QCUMbER).
\end{acknowledgments}

\end{document}